# A superradiant laser based on two-photon Raman transition of caesium atoms

**Pengfei Liu, Yu Peng***

*School of Physics, Beijing Institute of Technology, Beijing, 100081, P. R. China*

**Corresponding author:* pengyu@bit.edu.cn

We propose a superradiant laser based on two-photon Raman transition of caesium-133 atoms which collectively emit photons on an ultra narrow transition into the mode of a low Q resonator known as optical bad-cavity regime. The spin-spin correlation $\langle \hat{\sigma}_1^+ \hat{\sigma}_2^- \rangle_C$ which characterizes the collective effect is demonstrated. We theoretically predict that the optical radiation has an extremely narrow linewidth in the 98 (1) $*10^{-2}$ mHz range, smaller than the transition itself due to collective effects, and a power level of 7 (1)$*10^{-10}$ W is possible, which can provide a possible new way to realize an optical clock with a millihertz linewidth.

Narrow linewidth lasers are highly desirable for applications such as optical atomic clock work [1-2], gravitational wave detection, cavity quantum electrodynamics [3], quantum optomechanics [4]and precision tests of relativity [5].The best spectral purity of a laser is reached using an external optical resonator, an arrangement of two highly reflective mirrors that allows light to bounce back and forth between them many times. The laser linewidth is limited to 125 mHz because of integrated phase drift, which is induced by vibration of cavity mirrors due to thermal noise unable to be eliminated [6-7].

Recent researches indicate that laser using atom's ultra narrow optical transition operating in superradiant mode can produce a much purer spectra whose linewidth can be several orders of magnitude narrower than lasers in traditional laser regime. The collective atomic dipoles can be isolated from the external world by a factor of about ten thousand due to being operated with low intracavity photon number. And the output light holds a frequency linewidth only one in ten thousand of the quantum linewidth limit usually applied to lasers working in good-cavity regime. The optical radiation has an extremely narrow linewidth in the 119 (1)*$10^{-2}$ mHz range, and a power level of 3 (1)*$10^{-10}$ W according to [8].

Here, we propose a superradiant laser in which more than a million caesium-133 atomic dipoles are synchronized by 20 photons on average in the optical bad-cavity regime, and theoretically predict that the optical radiation has an extremely narrow quantum linewidth limit in the 98 (1) *$10^{-2}$ mHz range, and a power level of 7 (1)*$10^{-10}$ W will be reached in our caesium-133 atomic dipoles model which is possible to be better than rubidium-87.

The superradiant laser source in which spontaneous synchronization of nearly a million caesium-133 atomic dipoles are continuously sustained by average 20 photons inside the optical cavity as shown in Fig.1(a). A portion of atomic dipoles in gain medium reach synchronization under the influence of weak intracavity light field. They store the phase information, and nearly all the emitted photons escape the cavity rapidly, a phenomenon known as superradiance [8-9].

Two-photon processes have become one of the standard tools in atomic physics for exciting atoms to states whose energies are too high to achieve with a single photon, and also to states of the same parity that would normally be inaccessible. Especially a number of ultrahigh resolution spectroscopic techniques are based on two-photon processes. We consider two-photon Raman transition process as a model system in Fig. 1(b). In this case, the effective excited state, |e>=|$6^2\mathrm{S}_{1/2}$,F=2,$m_f = 0$>, and ground state, |g>=|$6^2\mathrm{S}_{1/2}$,F=1,$m_f$=0>, are hyperfine states and insensitive to magnetic field. The difference frequency between |e> and |g> is only 9.19 GHz. We impose a two photon stimulated Raman decay from |e>to |g> by applying a linearly polarized

894.6 nm dressing laser tuned 100 MHz higher than the intermediate state |i>, |i±>=| $6^2P_{1/2}$ ,F'=2, $m'_f$ =±1>. The transition $|\mathrm{e}>\rightarrow|\mathrm{i}>$ occurs by absorbing a photon at frequency $\omega_1$, and $|\mathrm{i}>\rightarrow|\mathrm{g}>$transition emitting a photon at frequency $\omega_2$. The frequencies are related by $\omega_2=\omega_1+\omega_{\mathrm{eg}}$. We then obtain for the transition probability $W_{e\rightarrow g}$ according to the reference [10-11].

$$W_{e\rightarrow g}=\frac{1}{(4\hbar^2)^2}\frac{|H_{gi,2}|^2|H_{ie,1}|^2}{(\omega_1-\omega_{ie})^2}\frac{(sin^2(\omega_0-\omega_1+\omega_2))t/2}{[(\omega_0-\omega_1+\omega_2)/2]^2} \tag{1}$$

Denoting the detuning of the intermediate state by $\Delta=\omega_1-\omega_{ie}$, more details are displayed in [10-11]. Cs atoms with dressing pump light from |e>=| $6^2\mathrm{S}_{1/2}$ ,F=2, $m_f$ =0> through intermediate states |i±>=|$6^2P_{1/2}$,F'=2,$m'_f$=±1> to |g>=|$6^2\mathrm{S}_{1/2}$,F=1,$m_f$=0> of the transmit path. Intermediate state to the ground state of atomic transition probabilities is 2.86× $10^7 s^{-1}$, lightly smaller than rubidium atoms, 3.6× $10^7 s^{-1}$. So we design the virtual state near |i> with $\Delta=\omega_1-\omega_{ie}$ =100 MHz, which will be helpful for transition probability according to formula (1).

The dressing laser is typically tuned 100 MHz to the blue of the |e>—|i±> atomic transition. There pumping lasers are polarized and tuned to the frequency between the ground states and the optical excited state |$5^2P_{3/2}$,F'=2> such that the single state dark to the repumping is |e>. The F1 repumper moves atoms primarily from the ground |F=1> state to the ground |F=2> state, and the F2 repumper pushes population to |e>, as the Clebsch-Gordan coefficient for the transition |$5^2S_{1/2}$,F'=2, $m'_f$=0> —>|$5^2P_{3/2}$, F'=2, $m'_f$=0> is zero, shown in Fig.1(c). Single-particle decay rate, $\gamma_{eg}$, from |e> to |g> is induced by exerting both repumping and dressing laser. And we can use a 852.3 nm light to repump atoms from |g> back to |e> at a proper rate *w* proportional to the single-particle decay rate to complete the whole circulation.

To understand the role of the collective effects, according to reference [12], we have the equation

$$0 = \langle \hat{\sigma}_1^+ \hat{\sigma}_2^- \rangle_C \begin{pmatrix} -\left(\gamma + \mathrm{w} + \frac{2}{T_2}\right) + \\ \frac{\mathrm{w}-\gamma}{(w+\gamma)} \mathrm{N}\gamma\mathrm{C} - 2\frac{(\mathrm{N}\gamma\mathrm{C})^2}{\gamma+\mathrm{w}} \langle \hat{\sigma}_1^+ \hat{\sigma}_2^- \rangle_C \end{pmatrix} \tag{2}$$

Where $\Gamma = \gamma + \mathrm{w} + \frac{2}{T_2}$ is the total decay rate of the atomic dipoles, the equation is built according to quantum model of the system and physically stable solution. The laser threshold is the repumping rate at which the gain $\frac{\mathrm{w}-\gamma}{(w+\gamma)} \mathrm{N}\gamma\mathrm{C}$ overcomes the losses $\Gamma$. In the limit $\Gamma/\mathrm{N}\gamma\mathrm{C} \rightarrow 0$, this condition turns into $\mathrm{w} > \gamma$. At this threshold the spin-spin correlations change sign to positive signifying the onset of collective behavior, shown in Fig. 2(a).

Interestingly, the spin-spin correlations change sign again at a larger pump rate above which the atoms return to normal non-collective emission. This upper threshold comes about because $\frac{\mathrm{w}-\gamma}{(w+\gamma)}$ eventually saturates at 1, while the pump induced noise grows with $w$. Setting $\frac{\mathrm{w}-\gamma}{(w+\gamma)} = 1$ and neglecting all atomic noise sources other than $w$, we find the maximum repumping rate $\mathrm{w}_{\mathrm{max}} = \mathrm{N}\gamma\mathrm{C}$. Above this threshold the pump noise destroys the coherences, shown in Fig. 2(b). Equation (2) also allows us to determine the maximum spin-spin correlation of $\langle \hat{\sigma}_1^+ \hat{\sigma}_2^- \rangle_C = 1/8$ which is obtained at the repumping rate $w_{opt} = \mathrm{N}\gamma\mathrm{C}/2$. At this pump rate the laser power reaches its maximum of $P_{\max} = \mathrm{h}\omega\mathrm{N}^2\gamma\mathrm{C}/8$.

The linewidth of a laser can be given by modified Schawlow-Townes full-width at half-maximum (FWHM) equation, $\Delta f_{ST}$

$$\Delta f_{ST} = \frac{1}{4\pi} \frac{\mathrm{h}f}{P_{out}} \left(\frac{2\gamma_{\perp}\kappa}{2\gamma_{\perp}+\kappa}\right)^2 \tag{3}$$

Here $P_{out}$ is the laser's output power, f is the oscillation frequency, κ is the cavity power decay rate due to mirror transmission alone and h is Planck constant. The transverse decoherence rate of the optical transition can be expressed as $\gamma_{\perp} = \frac{\gamma_{eg}}{2} + 1/T_2$, where $\gamma_{eg}$ is the single-particle decay rate from the upper state to the ground state and $1/T_2$ is induced by other atomic dephasing mechanisms such as spin dephasing.

If the cavity resonance frequency, $f_{cav}$ and atomic transition frequency, $f_{atomic}$ do not match, the system will oscillate at a weighted frequency

$$f = \frac{2\gamma_{\perp} f_{cav} + \kappa f_{atomic}}{2\gamma_{\perp} + \kappa} \tag{4}$$

The cavity resonance frequency $f_{cav}$ pulls the weighted frequency f from the atomic transition frequency by the level, $P = \frac{df}{df_{cav}} = 2\gamma_{\perp}/(2\gamma_{\perp} + \kappa) \approx 2\gamma_{\perp}/\kappa$ called the frequency pulling coefficient. We can also get the minimum of atom number $N = \frac{2}{T_2 \gamma C}$. Below this critical number, $\langle \hat{\sigma}_1^{+} \hat{\sigma}_2^{-} \rangle_C$ is never positive no matter how strong the repumping is and the collective dipoles hence will not be constructed. The critical particle number can be estimated from Eq. (2). Physically, this equation means that there must be enough atoms for the system to get into the strong coupling regime. The red points shown in all figures below represent corresponding value in our theoretical model. The scaling of that power with the number of atoms underlines the collective nature of the emission in Fig. 3, which shows that an output power of 7 (1)*$10^{-10}$ W is possible in our theoretical model, the number of atoms $N=10^6$ (the red point).

In Fig. 4(a), we demonstrate cavity pulling coefficient P around $4 \times 10^{-5}$ for a range of $\gamma_{eg}$ values. In the system, a bunch of $N=10^6$ $^{133}Cs$ atoms are confined to a low-finesse (F=300) optical cavity with cavity power decay rate $\kappa = 2\pi \times 25$ MHz, $T_2 \approx 0.00032$ s. As the decay rate $\gamma_{eg}$ decreases, the atomic dipole becomes more isolated from the mirrors, as shown by the frequency pulling coefficient $P$. In the bad-cavity, the frequency pulling coefficient $P \approx 2\gamma_{\perp}/\kappa \ll 1$ drastically reducing the impact of noise in the cavity frequency. This isolation of the oscillator from the environment is the key to abating the sensitivity of such a laser to thermal and technical noise. Output wavelength of the cesium superradiant laser system is pulled by the cavity resonance frequency (expressed in wavelength, ranging from 700 to 1100nm), $\frac{2\gamma_{\perp}}{\kappa}$=0.00004，0.01，0.1 respectively in Fig. 4(b). Figure 4(c) is a close look at the weighted wavelength changing with $\frac{2\gamma_{\perp}}{\kappa}$=0.00004.

In the optical bad-cavity, with limit of $2\gamma_{\perp} \ll \kappa$, the FWHM reduces to $\Delta f_{BST} = \gamma_{\perp}^2/(\pi\kappa M_c)$, shown in Fig.5(a). With $\frac{2\gamma_{\perp}}{\kappa} = 1 \times 10^{-5}$ to $1 \times 10^{-3}$, the excited state scattering rate, $\gamma_{eg} \approx 0.66 - 66s^{-1}$, is comparable to those atoms used on optical clock, and average intra-cavity photon number $M_C = 20$. Linewidth of the system slightly changes with the cavity frequency according to Shawlow-Townes equation with different value of $\frac{2\gamma_{\perp}}{\kappa}$ in Fig. 5(b). At 894.6 nm, the linewidth is only 98 (1) $*10^{-2}$ mHz, more than 10% less than rubidium laser (Fig. 5(c)).

In order to demonstrate the line shape of the power spectrum, we get the line-shape function by following theory. $b^{+}$ is the photon creation operator which satisfying:

$$b^{+} = e^{i\varphi}(r_0 + \rho)e^{i\Omega t} \quad (5)$$

Where $\varphi$ is the phase shift due to noise, $r_0$ is a classical number, and $\rho$ is density matrix operator presenting the fluctuation of the amplitude. Output frequency is $\Omega = \frac{\nu_c\chi+\gamma\omega}{\chi+\gamma}$, where $\chi$ is the linewidth of the cavity, and $\nu_c$ is the cavity's resonance frequency. $\gamma$ and $\omega$ are the self-decay linewidth and transition frequency of the atoms, respectively. Assuming $\rho \ll r_0$ and $e^{i\Omega t}$ can be separated, we can get $b^{+} \approx r_0 e^{i\varphi(t)}$ from (5). The correlation function is defined as $\langle b^+(t)b(0)\rangle$. And $\varphi_\mu$ is independent to each other, $\varphi(t)-\varphi(0)=\sum_\mu \varphi_\mu$.

$$\langle b^+(t)b(0)\rangle \approx r_0^2 \prod_\mu \langle e^{-i\varphi_\mu(t)}\rangle \approx r_0^2 \prod_\mu \left\{1+i\langle\varphi_\mu(t)\rangle-\frac{1}{2}\langle\varphi_\mu(t)^2\rangle\right\} \tag{6}$$

With $\langle\varphi_\mu\rangle=0$, $\langle\varphi_\mu\varphi_{\mu'}\rangle=0$ $(\varphi_\mu\neq\varphi_{\mu'})$, we can get $\sum_\mu\langle{\varphi_\mu}^2\rangle=\langle(\varphi(t)-\varphi(0))^2\rangle$. Combined with (5), we get:

$$\langle b^+(t)b(0)\rangle = r_0^2 e^{-\langle(\varphi(t)-\varphi(0))^2\rangle\cdot\frac{1}{2}} \tag{7}$$

Assuming $\Delta\omega=\frac{1}{2t}\langle(\varphi(t)-\varphi(0))^2\rangle$, $(t>\frac{1}{\chi+\gamma})$, we get

$$\langle b^+(t)b(0)\rangle = r_0^2 e^{-\Delta\omega t} \tag{8}$$

Then multiplying the formula above by a factor $e^{i\Omega t}$ and by Fourier transformation we get the spectrum, shown in Fig.6. Squaring and normalizing it, we could get the line-shape function.

$$g_N(\nu,\Omega)=\frac{2\Delta\omega}{\Delta\omega^2+4\pi^2(\nu-\Omega)^2} \tag{9}$$

$\Omega$ is the center frequency of the oscillation, and linewidth induced by the phase noise is $\Delta\omega/2\pi$. Linewidth deceases as the output power increasing, which is just the opposite case in good cavity regime (green line) which is shown in Figure 7(a). Compared to Rb laser, Cs laser has narrower linewidth with the same output power. Figure 7(b) shows that an output power of $7*10^{-10}$ W is possible with the linewidth of 0.99 mHz.

Because of the narrow linewidth, it could be applied in frequency standard. The layout for the new optical clock scheme is shown in Fig. 8. Repumping laser, at 852.3 nm, is constructed by an extended cavity laser diode (ECDL) resulting in ~ MHz linewidth. Dressing laser, at 894.6 nm, is stabilized by ultra-stable cavity resulting in ~1 Hz linewidth. Both lasers pass and couple with the $^{133}$Cs atoms, which are trapped in the magic optical lattice. In ref.8, the present linewidth is limited by atomic population noise. In our proposed active optical clocks, aiming to reduce atom loss, we think that the atom loss mechanism, while important, is not essential to the underlying physics and can be eliminated in future work by using higher-dimensional lattices, for example, 3-dimensional lattices (green arrow), $^{133}$Cs atoms that are trapped in magic optical lattice at 852 nm [13-14]. Moreover, with lower repumping light intensity, narrower repumping linewidth, about 100 Hz, we think it's possible that the superradiant laser emits photons into optical bad-cavity regime, from which an optical frequency standard at 894.6 nm with less than 1 mHz linewidth is output.

In summary, we predict that the optical radiation has an extremely narrow linewidth in the 98 (1) $*10^{-2}$ mHz range, and a power level of 7 (1)$*10^{-10}$ W, which is sufficient for phase locking a slave optical oscillator. We choose cesium system for its long wavelength transition of cesium-133. The frequency pulling coefficient becomes drastically reducing the impact of noise in the cavity frequency. This isolation of the oscillator from the environment is the key to reducing the sensitivity of such a laser to thermal and technical noise. Although the frequency pulling coefficient becomes drastically, the linewidth of oscillator can be impacted by its wavelength for some level. Therefore, we looked for all the alkali metal atoms, and find the transition probability of atomic ground state to the first excited state transitions based on Cs is large and the wavelength (894.6 nm) is longer than the rubidium (795 nm). It means that the linewidth limit becomes smaller that rubidium-87 in same conditions, and a superradiant laser in caesium-133 atomic dipoles theoretically has an extremely narrower quantum linewidth limit than rubidium system. Moreover, in order to reduce the frequency pulling coefficient further, we design optical bad-cavity regime with lower finesse (F=300), and the linewidth will be narrower.

For more power, we propose the system with more atoms, shown in Fig.3, and average 20

photons inside the optical cavity are maintained. It will guarantee the power with a level of 7 (1)*$10^{-10}$ W, which is important in practical applications. This mechanism can provide a possible new way to realize an optical clock with a millihertz linewidth, which would make atomic clocks more stable, since mHz linewidth is far beyond the present coherence time realized by passive optical clocks.

This work is supported by Basic Research Funds from Beijing Institute of Technology (Grant No. 20121842004).

**Figure captions:**

FIG.1. (Color online) Diagram of Cs superradiant laser. (a)The coherence is stored by the collective effect, and the intra cavity photons extract phase information at a suitable rate. That means the amplitude and phase are mostly stored in the atomic medium, and the laser frequency depends only very weakly on the distance between the mirrors. Pink and yellow balls represent photons and atoms respectively. (b)Two-photon Raman transition of caesium atoms. Dressing the metastable ground state |e> with a laser (red line) to induce a spontaneous two-photon Raman transition to |g> (blue line), with tunable transition rate $\gamma_{eg}$. The repumping laser move atoms from state |g> back to |e> in a proper rate and thus build a circle. (c) Repumping transitions. The energy level diagram for the repumping beam F2(blue) and F1(green).The repumping dark state is labeled with a red line.

Fig. 2. (Color online)Spin-spin correlation, $\langle\hat{\sigma}_1^{+}\hat{\sigma}_2^{-}\rangle_C$, as a function of the repumping rate, w, and decay rates, $\gamma_{eg}$ with fixed atom number N= $10^6$. With larger spin-spin correlation, there will be a stronger and more independent output as well as narrower linewidth. (a) Three dimensions graph. (b) Contour map.

FIG.3. (Color online)Output power as a function of atom number N. $\gamma = 17s^{-1}$,$C = 7.5 \times 10^{-3}$.

FIG.4. (Color online) (a) As the decay rate $\gamma_{eg}$ increases, the atomic dipoles become more dependent on the cavity mirrors, shown by the frequency pulling coefficient $P$. (b)Wavelength of the cesium superradiant laser system is pulled by the cavity resonance frequency (expressed in wavelength ranging from 700 to 1100 nm), $\frac{2\gamma_\perp}{\kappa}$=0.00004，0.01，0.1 respectively with fixed $\gamma_\perp$. (c) Close look of the weighted wavelength as a function of resonance wavelength of cavity ($\frac{2\gamma_\perp}{\kappa}$=0.0004).

FIG.5. (Color online) (a) Linewidth increases almost linearly as finesse goes from 14 to 400. (b) Linewidth of the system slightly changes with the cavity frequency according to Shawlow-Townes equation, $\frac{2\gamma_\perp}{\kappa}$=0.00004，0.01，0.1 respectively with fixed $\gamma_\perp$. (c) Close look of the FWHM as a function of wavelength of cavity ($\frac{2\gamma_\perp}{\kappa}$=0.00004). At 894.6 nm, the linewidth is only 0.99 mHz.

FIG.6. (Color online) The line shape of the power spectrum. The power spectrum has a Lorentzian shape with linewidth of 0.99 mHz.

FIG.7. (Color online) Relation between linewidth of laser and the output power with fixed $\kappa$, $f_{cav}$ and N. (a) Comparison between Cs superradiant laser (blue) and He-Ne laser in good cavity regime (green). (b) Cs superradiant laser alone.

FIG.8. (color online) Experimental layout. Two lasers are both constructed by extended cavity laser diodes, in which dressing laser is stablized by an ultra-stable cavity. These two lasers interact with $^{133}$Cs atoms that are trapped in magic optical lattice at 852 nm. The superradiant laser emits photons into optical bad-cavity regime, from which an optical frequency standard at 894.6 nm with ~1 mHz linewidth is output.

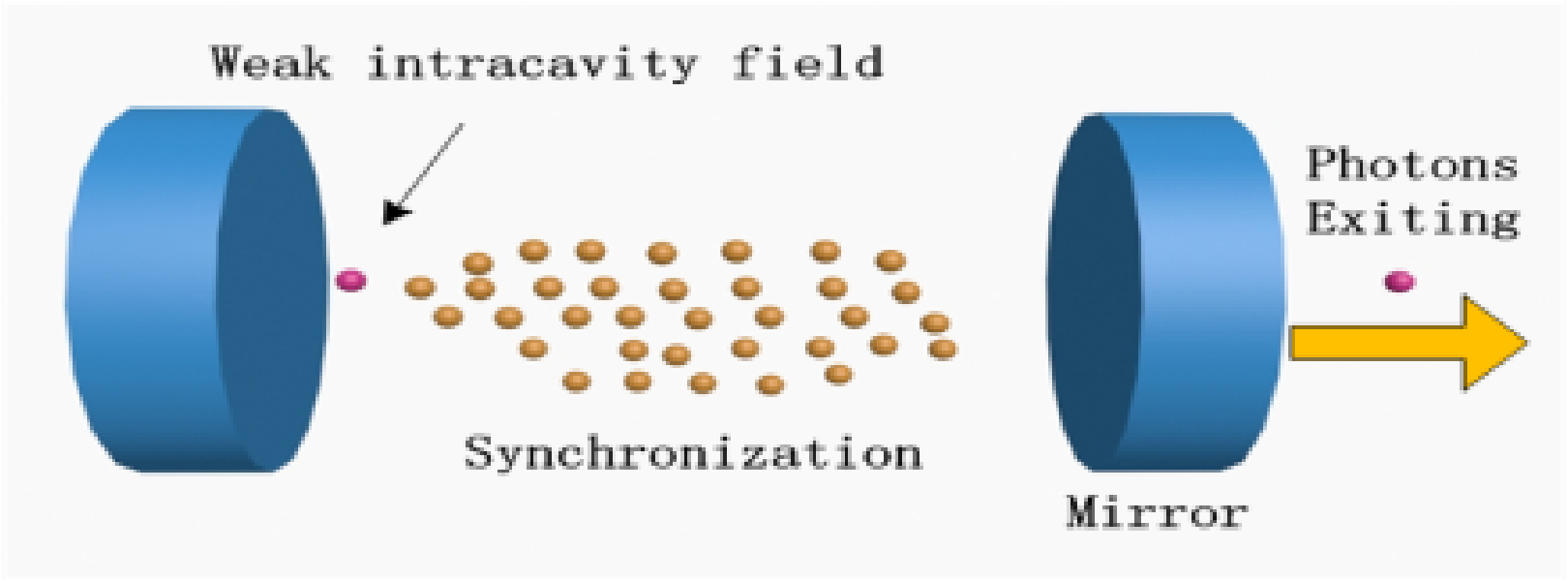
Weak intracavity field
Photons
Exiting
Synchronization
Mirror

(a)

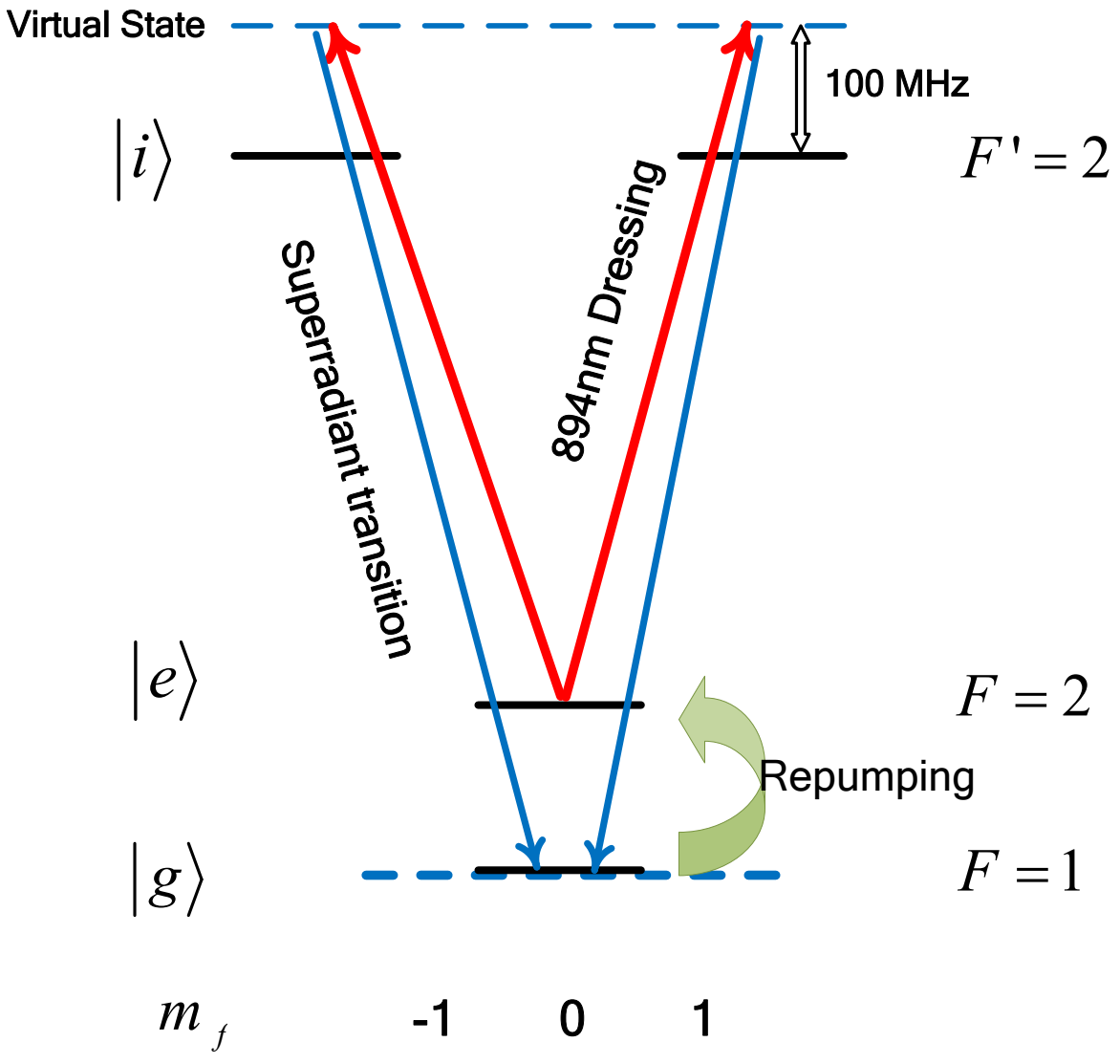

(b)

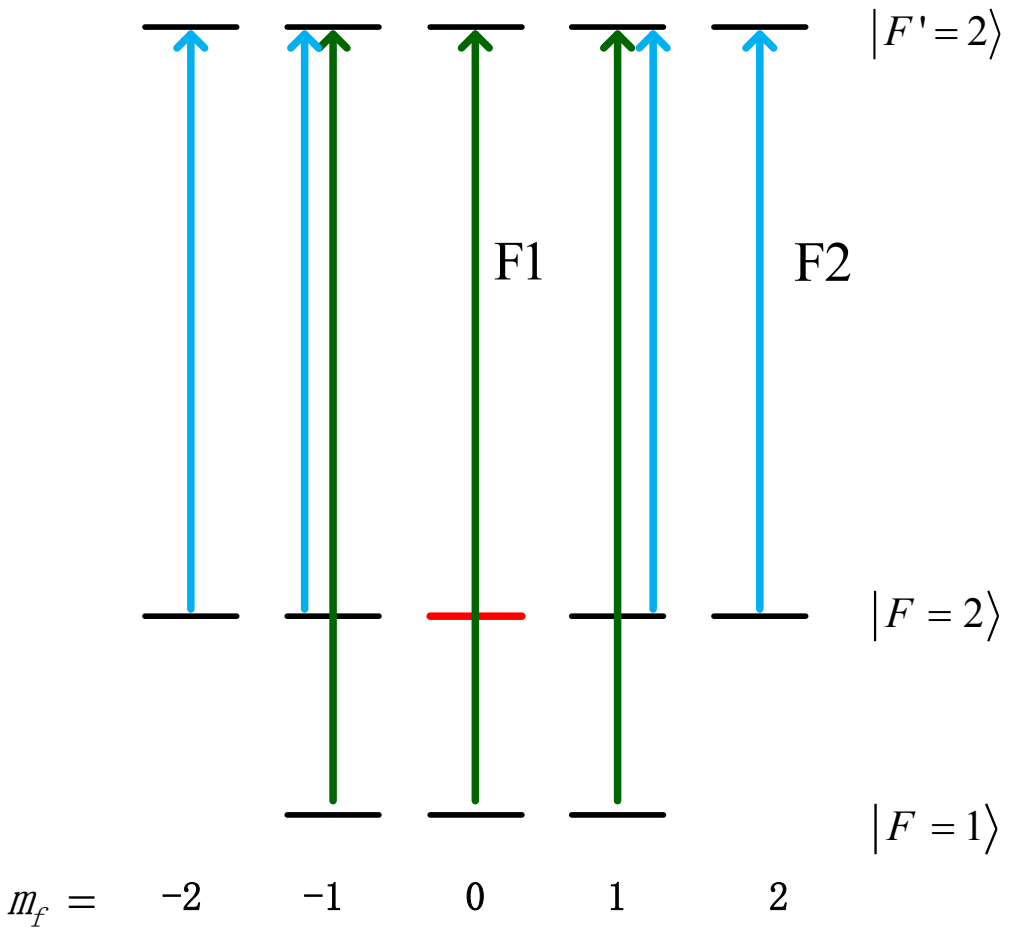

(c)

Fig. 1.

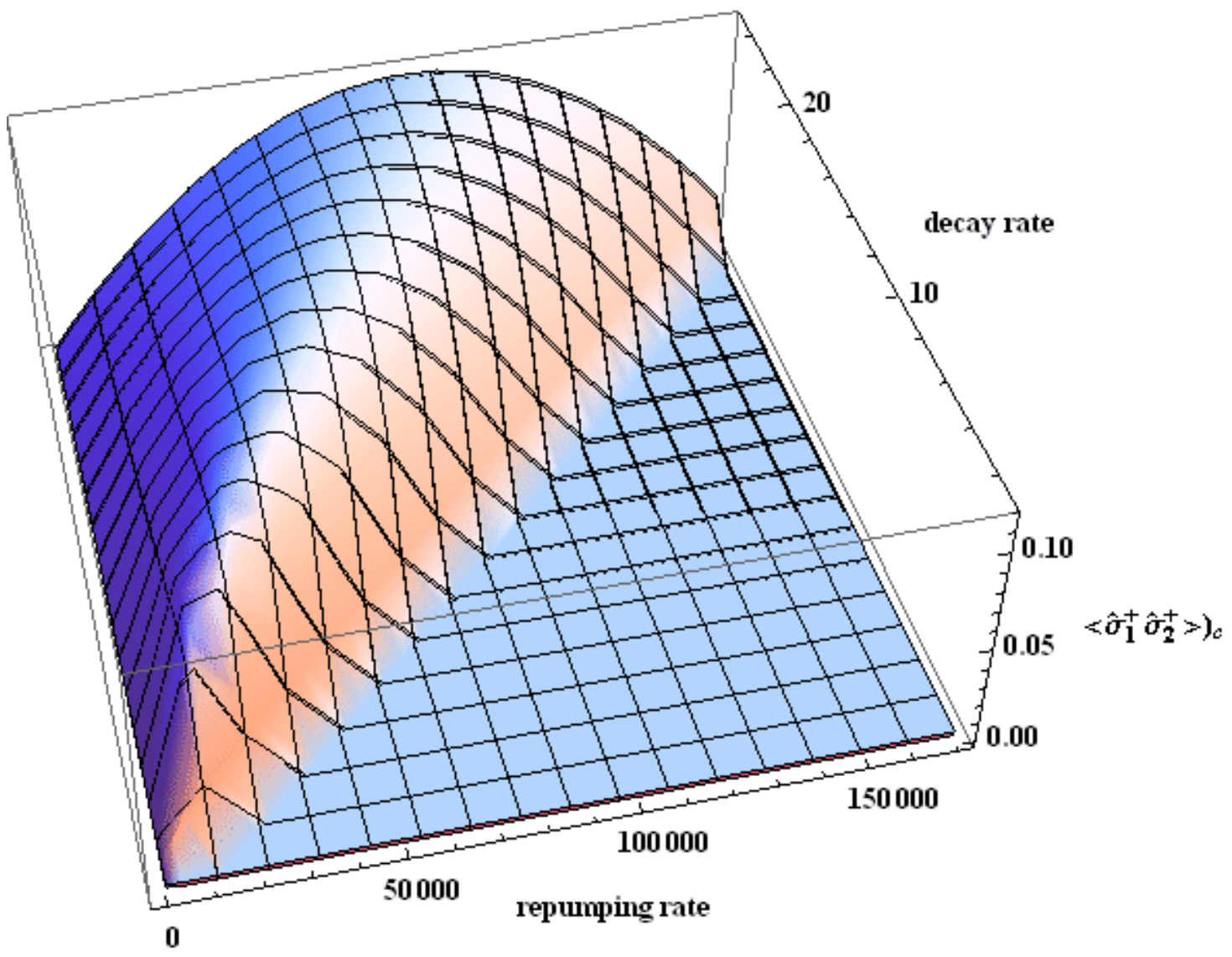

(a)

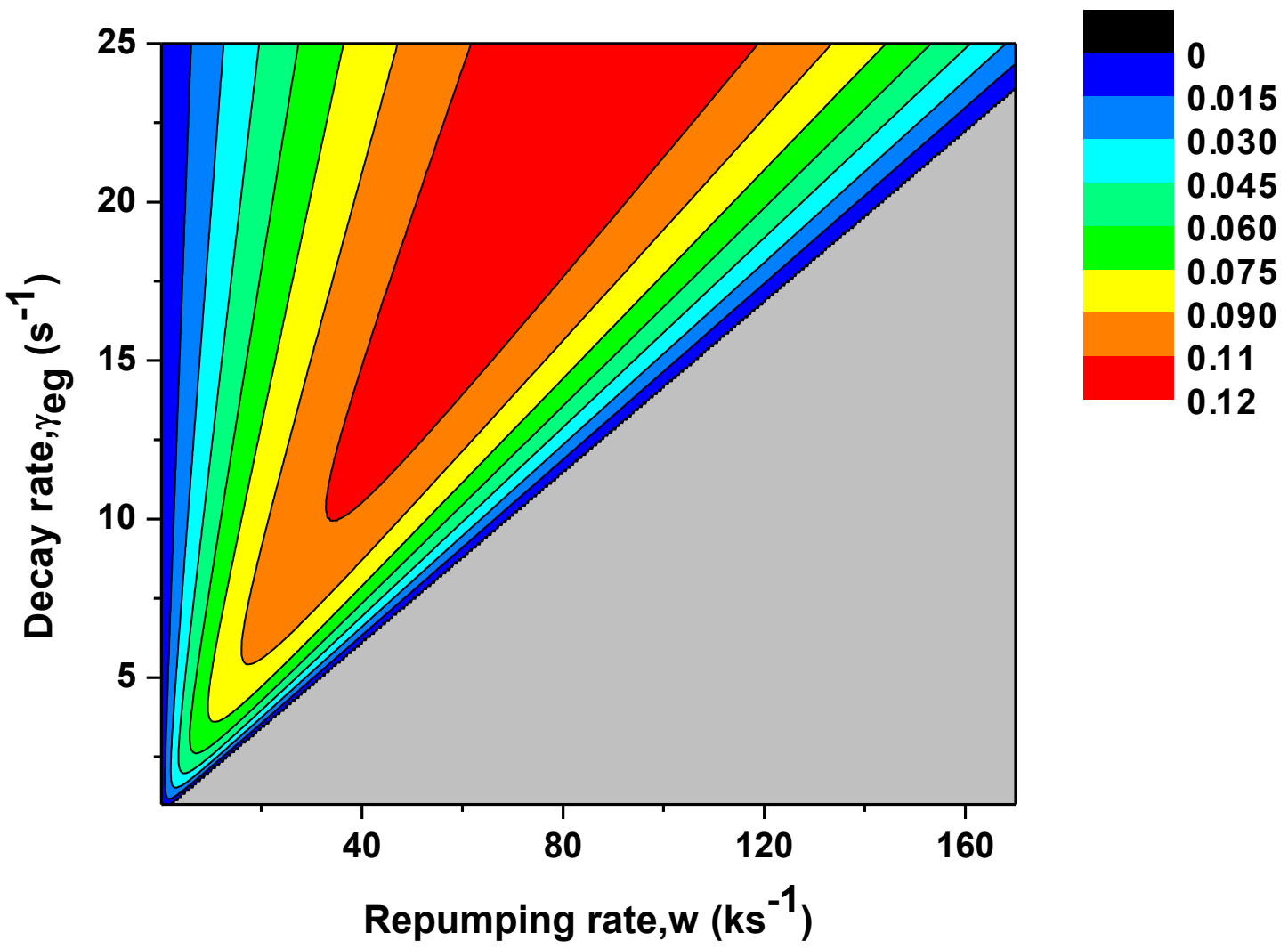

(b)

Fig. 2.

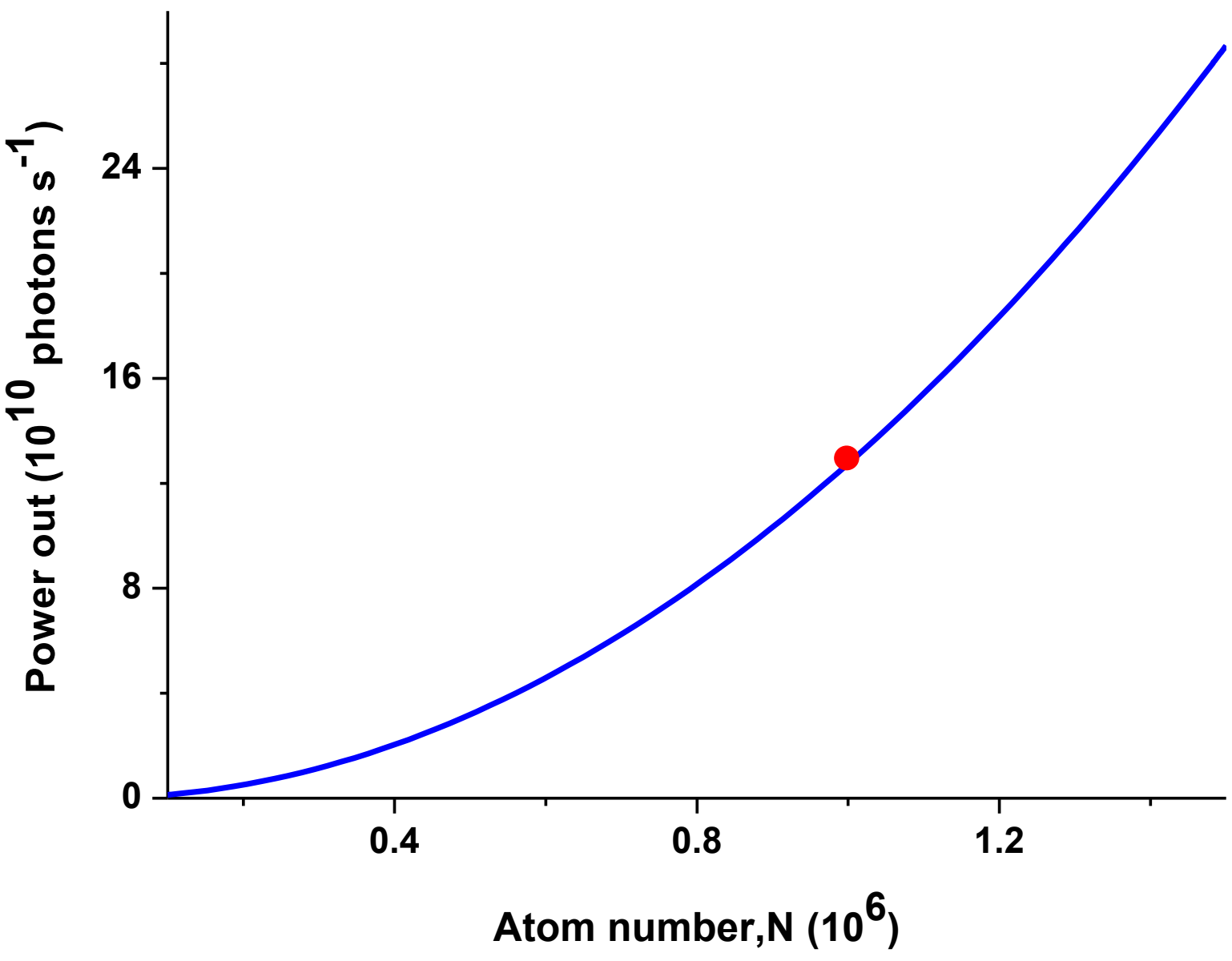

Fig. 3.

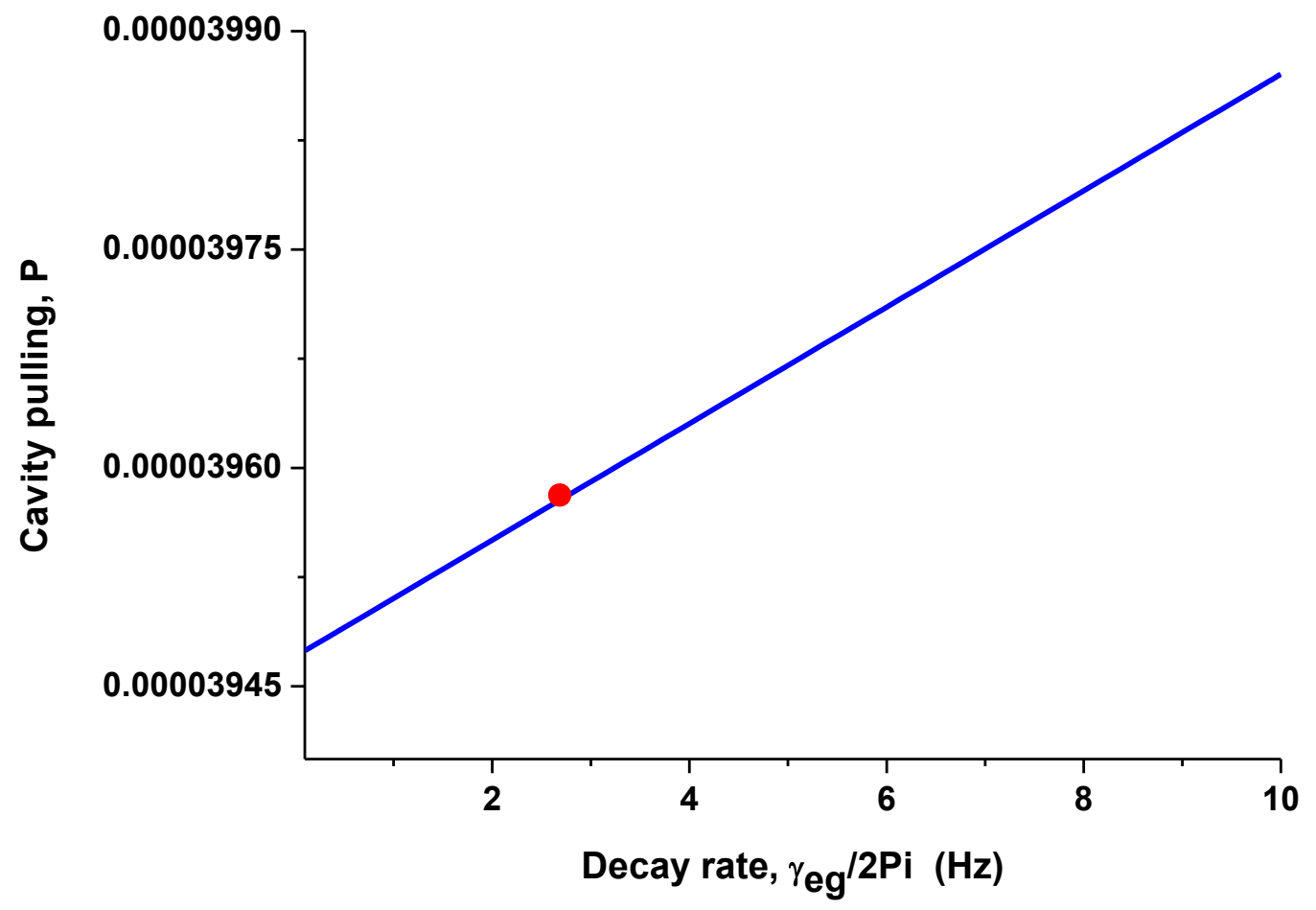

(a)

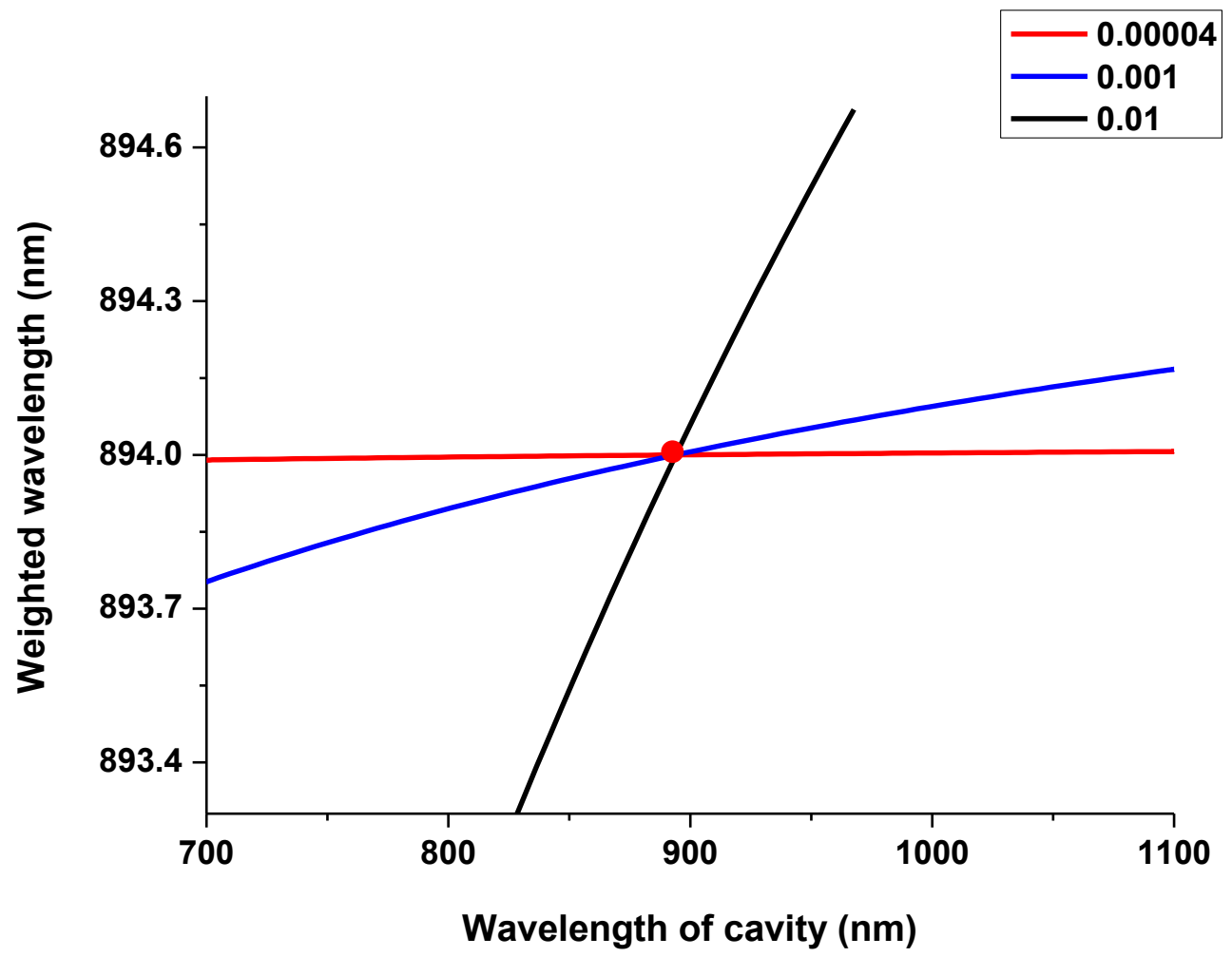

(b)

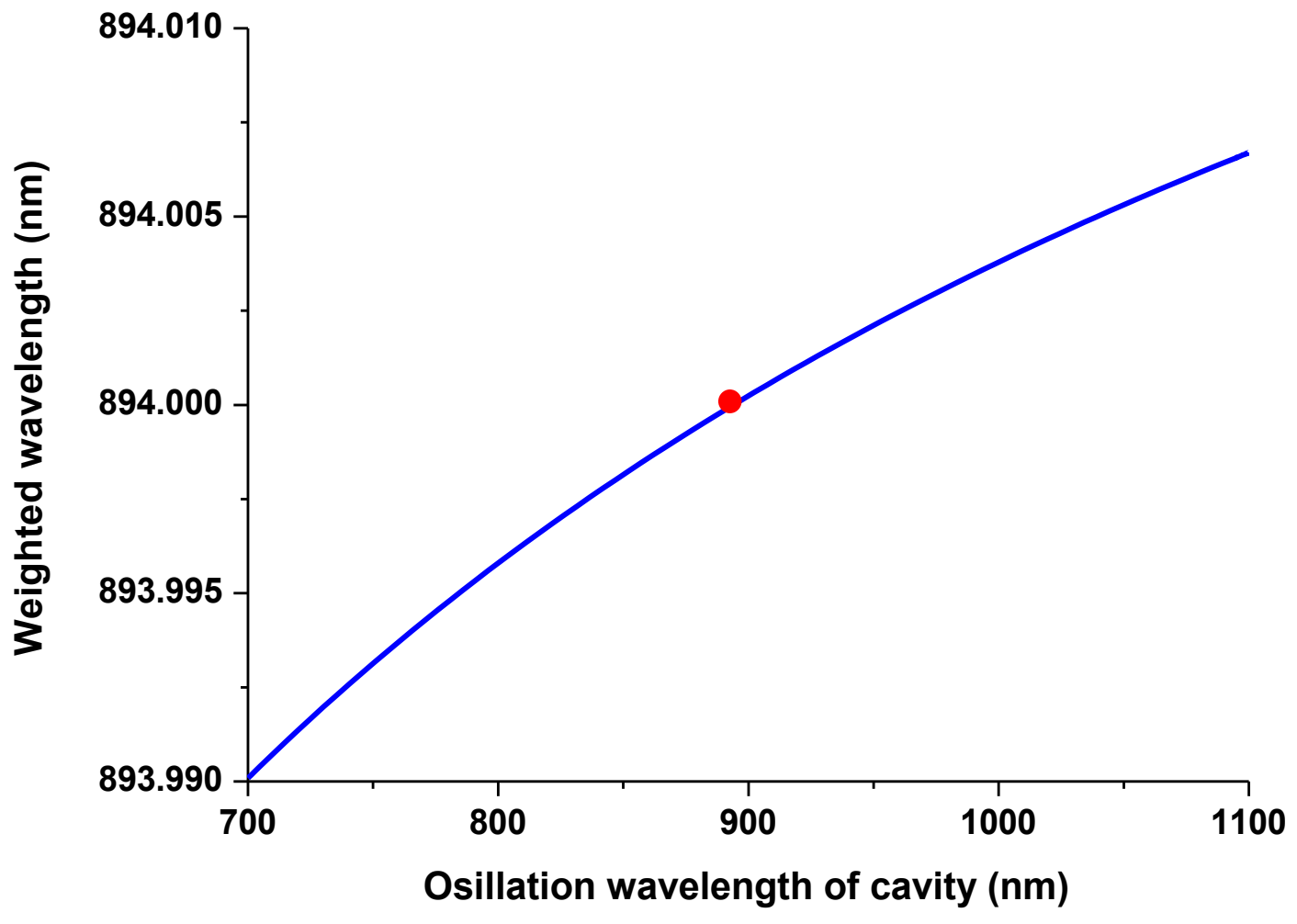

(c)

Fig. 4.

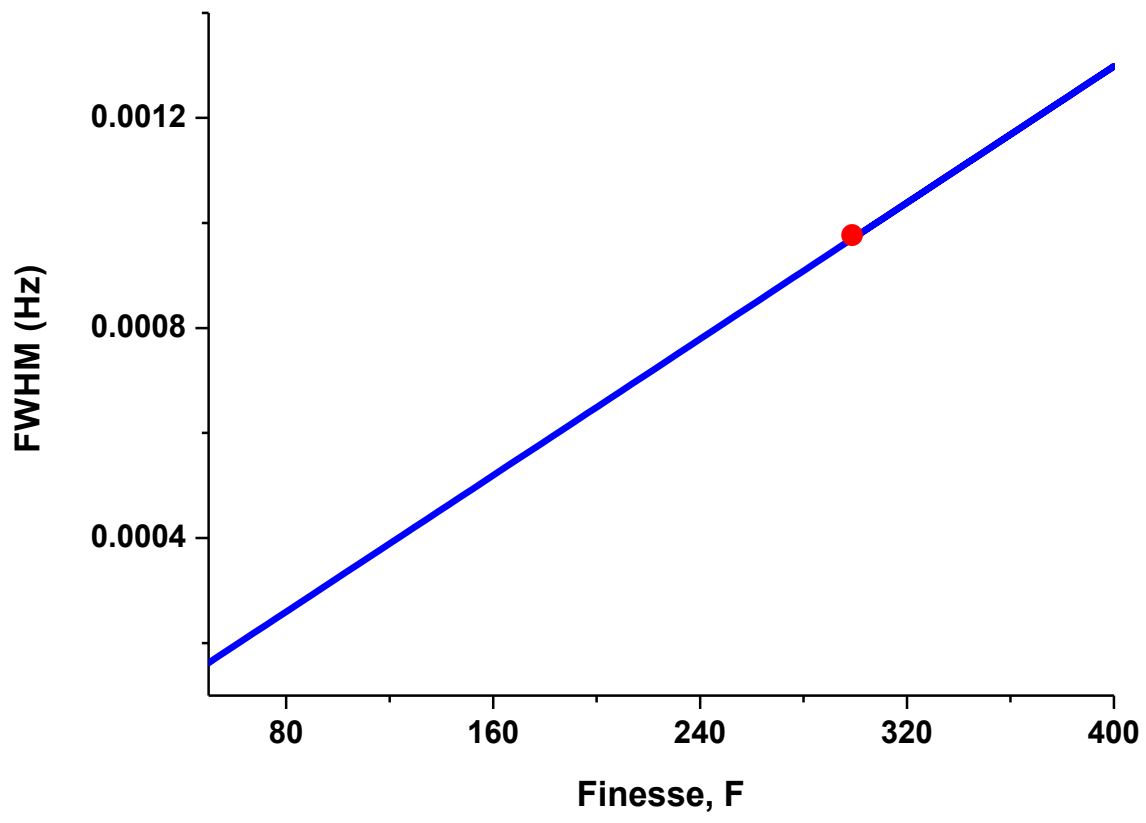

(a)

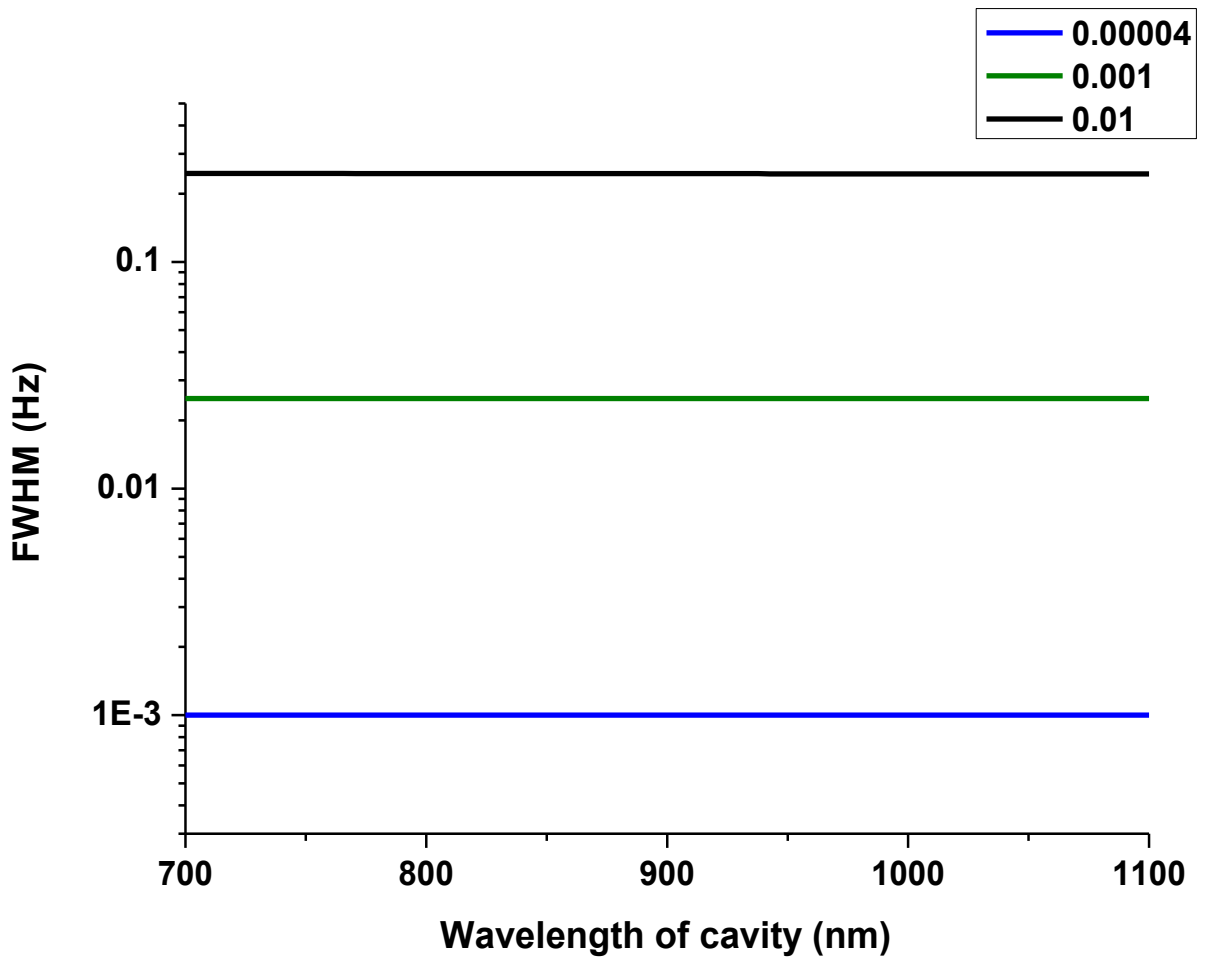

(b)

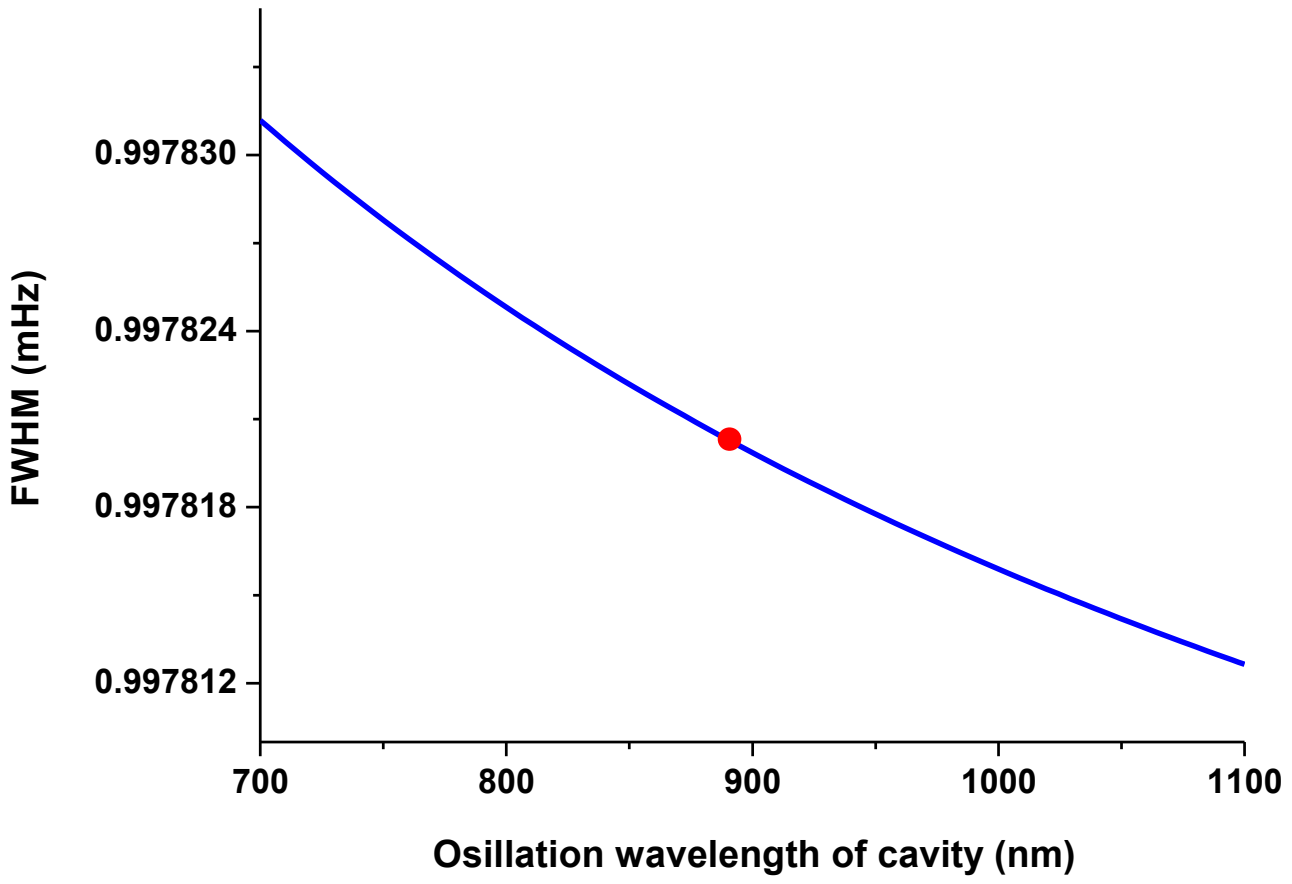

(c)

Fig. 5.

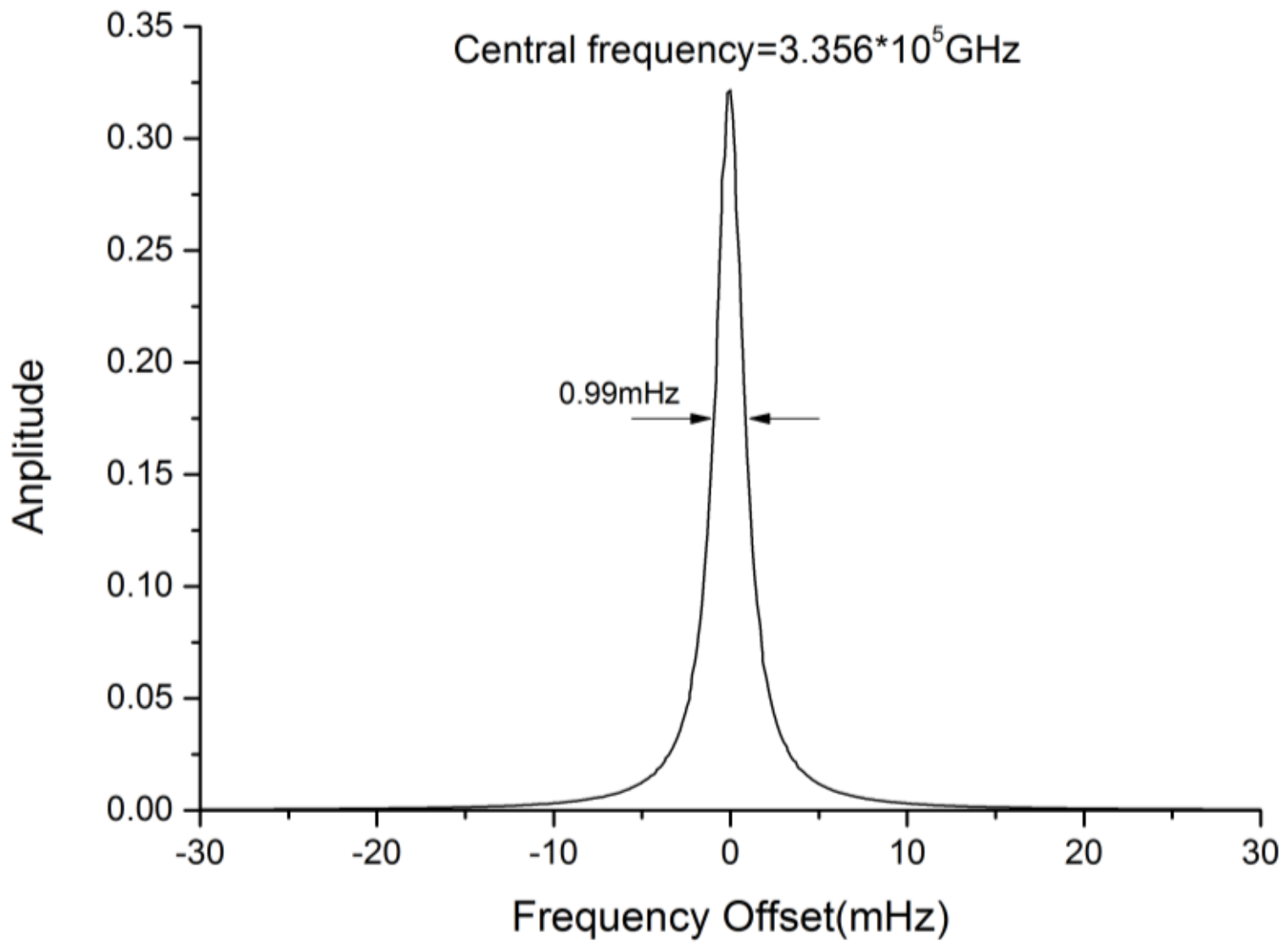
Central frequency=3.356*10^5GHz
0.99mHz
Anplitude
Frequency Offset(mHz)

Fig. 6.

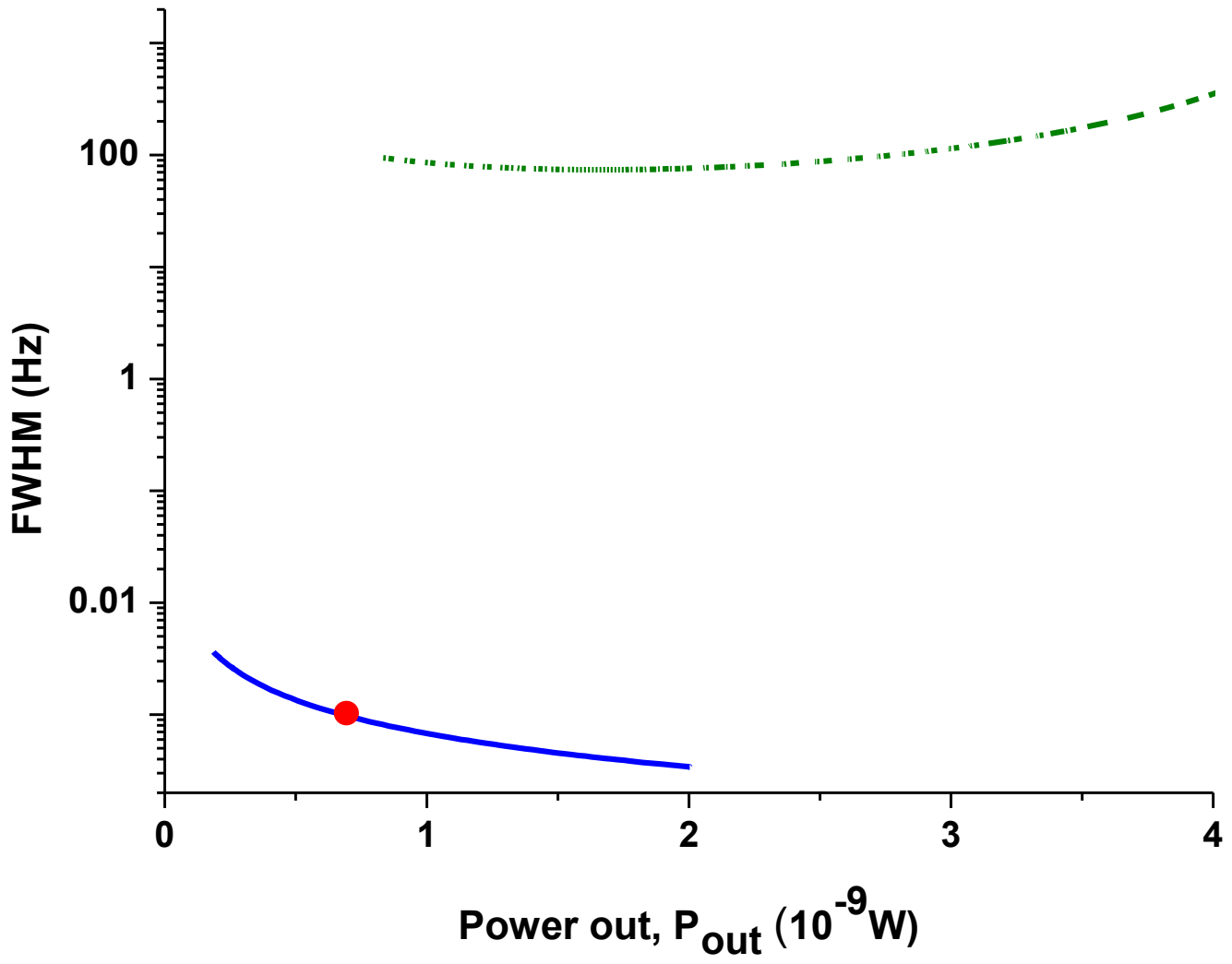
FWHM (Hz)
Power out, P_out (10^-9W)

(a)

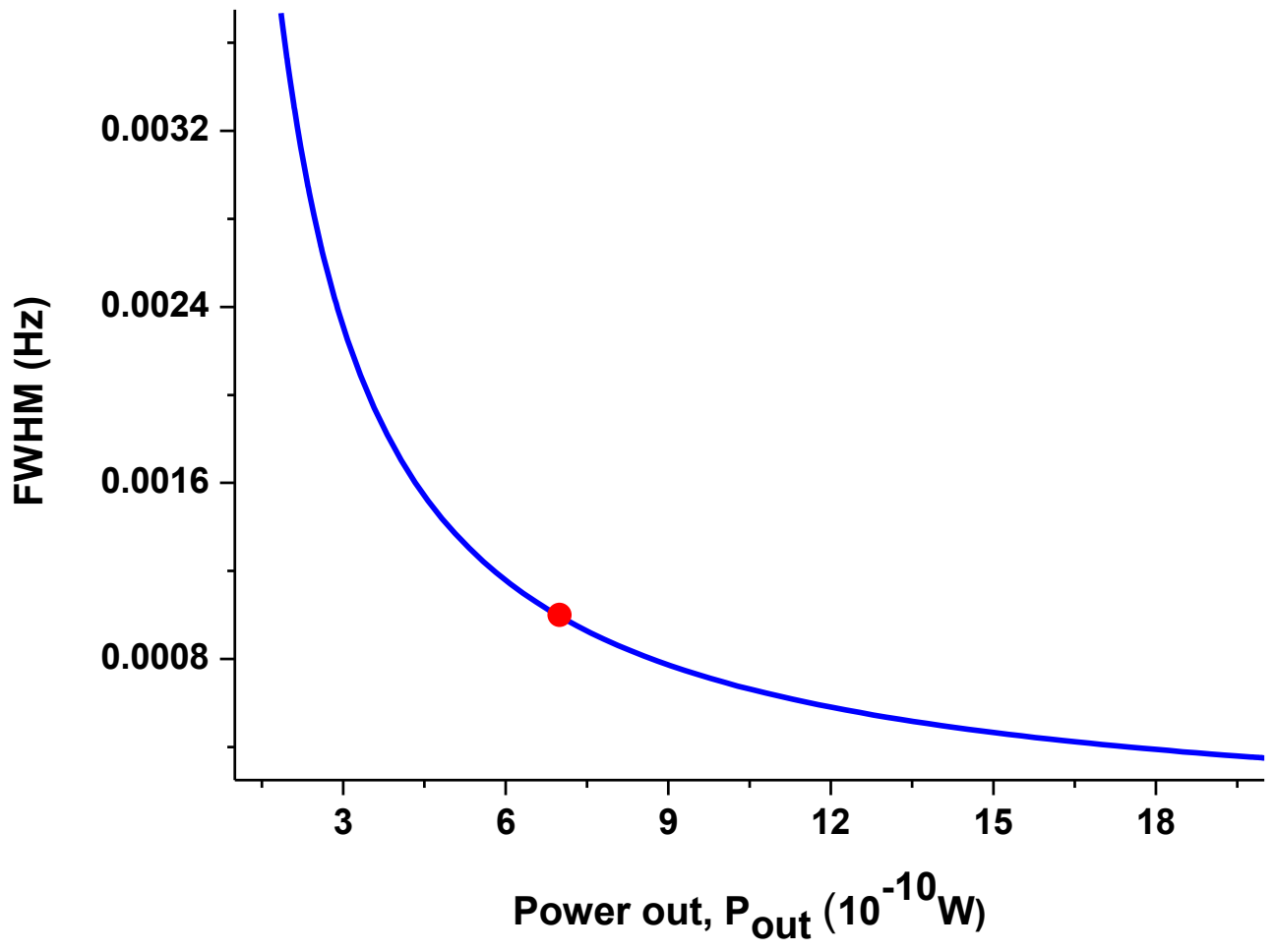

(b)

Fig.7.

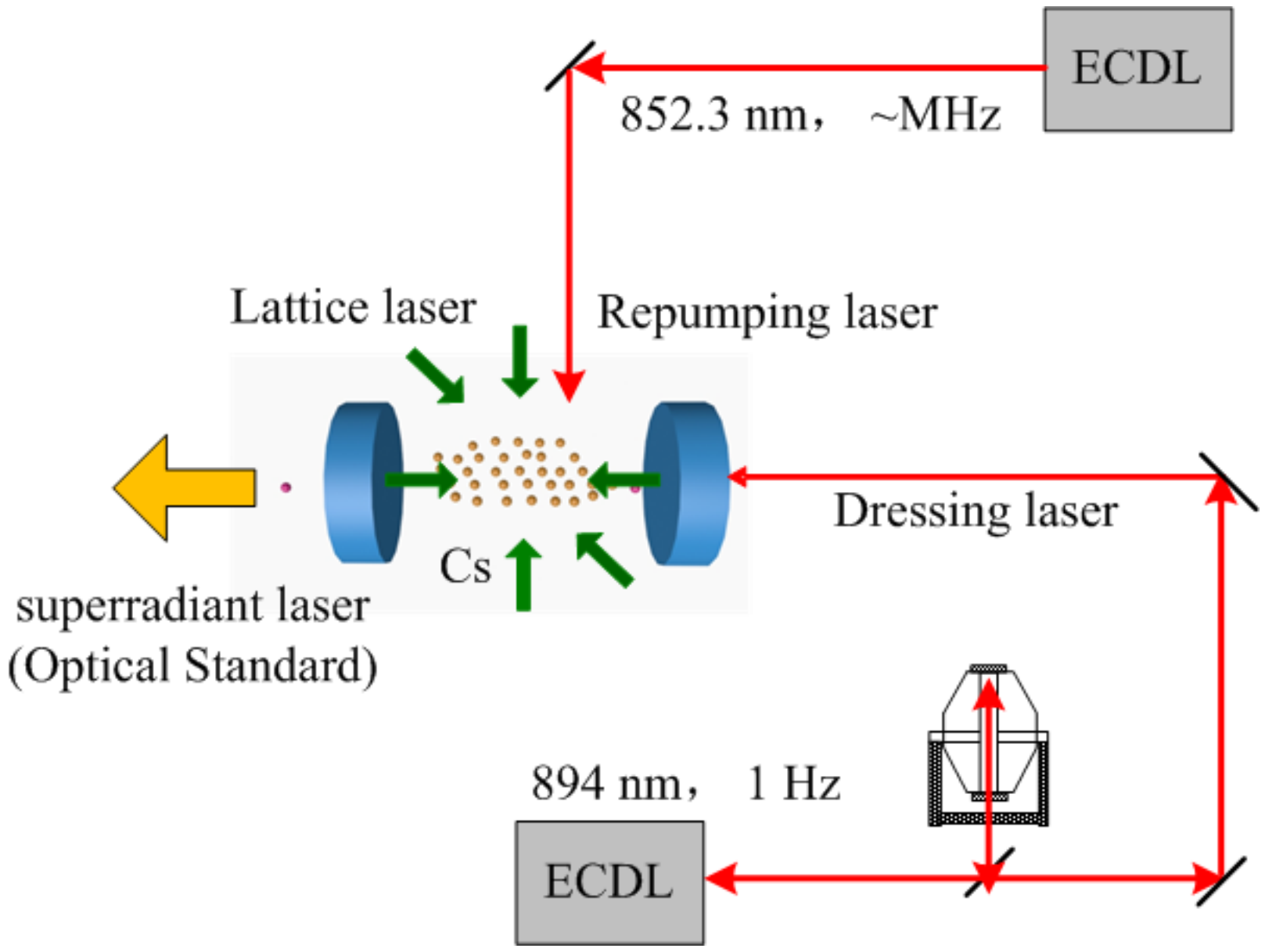

Fig.8.